\documentclass[
    amsmath,
    amssymb,
    aps,
    twocolumn,
    prl,
    showpacs,
    showkeys,
    preprintnumbers,
    superscriptaddress,
    nobibnotes,
    floatfix,
    longbibliography,
    notitlepage,
    nofootinbib
]
{revtex4-2}

\usepackage{amsmath}
\usepackage{amsfonts}
\usepackage{amssymb}
\usepackage{graphicx}
\usepackage{lipsum}
\usepackage[colorlinks=true,allcolors=blue]{hyperref}
\usepackage{relsize}
\usepackage{textcomp}
\usepackage{gensymb}
\usepackage{fontawesome}
\usepackage{siunitx}
\usepackage[capitalize]{cleveref}
\usepackage[dvipsnames,table]{xcolor}
\usepackage[toc,page]{appendix}
\usepackage[section]{placeins}
\usepackage{lineno}

\newcommand{\numu}{$\nu_{\mu}$}
\newcommand{\charon}{\texttt{\raisebox{0.82\depth}{$\chi$}aro$\nu$}}
\newcommand{\nusquids}{\texttt{nuSQuIDS}}

\newcommand{\lh}{\mathcal{L}}

\newcounter{starredsection}%

\makeatletter
\let\latex@@section\section

\AtBeginDocument{%
  \RenewDocumentCommand{\subsection}{som}{%
    \IfBooleanTF{#1}{%
      \refstepcounter{starredsection}%
      \latex@@section*{#3}%
    }{%
      \IfValueTF{#2}{%
        \latex@@section[#2]{#3}%
      }{%
        \latex@@section{#3}%
      }%
    }%
  }%
}
\makeatother

\crefname{starredsection}{App.}{Apps.}
\crefname{starredsection}{App.}{Apps.}

\begin{document}

\title{Search for High-Energy Neutrinos From the Sun Using Ten Years of IceCube Data}

\affiliation{III. Physikalisches Institut, RWTH Aachen University, D-52056 Aachen, Germany}
\affiliation{Department of Physics, University of Adelaide, Adelaide, 5005, Australia}
\affiliation{Dept. of Physics and Astronomy, University of Alaska Anchorage, 3211 Providence Dr., Anchorage, AK 99508, USA}
\affiliation{School of Physics and Center for Relativistic Astrophysics, Georgia Institute of Technology, Atlanta, GA 30332, USA}
\affiliation{Dept. of Physics, Southern University, Baton Rouge, LA 70813, USA}
\affiliation{Dept. of Physics, University of California, Berkeley, CA 94720, USA}
\affiliation{Lawrence Berkeley National Laboratory, Berkeley, CA 94720, USA}
\affiliation{Institut f{\"u}r Physik, Humboldt-Universit{\"a}t zu Berlin, D-12489 Berlin, Germany}
\affiliation{Fakult{\"a}t f{\"u}r Physik {\&} Astronomie, Ruhr-Universit{\"a}t Bochum, D-44780 Bochum, Germany}
\affiliation{Universit{\'e} Libre de Bruxelles, Science Faculty CP230, B-1050 Brussels, Belgium}
\affiliation{Vrije Universiteit Brussel (VUB), Dienst ELEM, B-1050 Brussels, Belgium}
\affiliation{Dept. of Physics, Simon Fraser University, Burnaby, BC V5A 1S6, Canada}
\affiliation{Department of Physics and Laboratory for Particle Physics and Cosmology, Harvard University, Cambridge, MA 02138, USA}
\affiliation{Dept. of Physics, Massachusetts Institute of Technology, Cambridge, MA 02139, USA}
\affiliation{Dept. of Physics and The International Center for Hadron Astrophysics, Chiba University, Chiba 263-8522, Japan}
\affiliation{Department of Physics, Loyola University Chicago, Chicago, IL 60660, USA}
\affiliation{Dept. of Physics and Astronomy, University of Canterbury, Private Bag 4800, Christchurch, New Zealand}
\affiliation{Dept. of Physics, University of Maryland, College Park, MD 20742, USA}
\affiliation{Dept. of Astronomy, Ohio State University, Columbus, OH 43210, USA}
\affiliation{Dept. of Physics and Center for Cosmology and Astro-Particle Physics, Ohio State University, Columbus, OH 43210, USA}
\affiliation{Niels Bohr Institute, University of Copenhagen, DK-2100 Copenhagen, Denmark}
\affiliation{Dept. of Physics, TU Dortmund University, D-44221 Dortmund, Germany}
\affiliation{Dept. of Physics and Astronomy, Michigan State University, East Lansing, MI 48824, USA}
\affiliation{Dept. of Physics, University of Alberta, Edmonton, Alberta, T6G 2E1, Canada}
\affiliation{Erlangen Centre for Astroparticle Physics, Friedrich-Alexander-Universit{\"a}t Erlangen-N{\"u}rnberg, D-91058 Erlangen, Germany}
\affiliation{Physik-department, Technische Universit{\"a}t M{\"u}nchen, D-85748 Garching, Germany}
\affiliation{D{\'e}partement de physique nucl{\'e}aire et corpusculaire, Universit{\'e} de Gen{\`e}ve, CH-1211 Gen{\`e}ve, Switzerland}
\affiliation{Dept. of Physics and Astronomy, University of Gent, B-9000 Gent, Belgium}
\affiliation{Dept. of Physics and Astronomy, University of California, Irvine, CA 92697, USA}
\affiliation{Karlsruhe Institute of Technology, Institute for Astroparticle Physics, D-76021 Karlsruhe, Germany}
\affiliation{Karlsruhe Institute of Technology, Institute of Experimental Particle Physics, D-76021 Karlsruhe, Germany}
\affiliation{Dept. of Physics, Engineering Physics, and Astronomy, Queen's University, Kingston, ON K7L 3N6, Canada}
\affiliation{Department of Physics {\&} Astronomy, University of Nevada, Las Vegas, NV 89154, USA}
\affiliation{Nevada Center for Astrophysics, University of Nevada, Las Vegas, NV 89154, USA}
\affiliation{Dept. of Physics and Astronomy, University of Kansas, Lawrence, KS 66045, USA}
\affiliation{Centre for Cosmology, Particle Physics and Phenomenology - CP3, Universit{\'e} catholique de Louvain, Louvain-la-Neuve, Belgium}
\affiliation{Department of Physics, Mercer University, Macon, GA 31207-0001, USA}
\affiliation{Dept. of Astronomy, University of Wisconsin{\textemdash}Madison, Madison, WI 53706, USA}
\affiliation{Dept. of Physics and Wisconsin IceCube Particle Astrophysics Center, University of Wisconsin{\textemdash}Madison, Madison, WI 53706, USA}
\affiliation{Institute of Physics, University of Mainz, Staudinger Weg 7, D-55099 Mainz, Germany}
\affiliation{Department of Physics, Marquette University, Milwaukee, WI 53201, USA}
\affiliation{Institut f{\"u}r Kernphysik, Universit{\"a}t M{\"u}nster, D-48149 M{\"u}nster, Germany}
\affiliation{Bartol Research Institute and Dept. of Physics and Astronomy, University of Delaware, Newark, DE 19716, USA}
\affiliation{Dept. of Physics, Yale University, New Haven, CT 06520, USA}
\affiliation{Columbia Astrophysics and Nevis Laboratories, Columbia University, New York, NY 10027, USA}
\affiliation{Dept. of Physics, University of Oxford, Parks Road, Oxford OX1 3PU, United Kingdom}
\affiliation{Dipartimento di Fisica e Astronomia Galileo Galilei, Universit{\`a} Degli Studi di Padova, I-35122 Padova PD, Italy}
\affiliation{Dept. of Physics, Drexel University, 3141 Chestnut Street, Philadelphia, PA 19104, USA}
\affiliation{Physics Department, South Dakota School of Mines and Technology, Rapid City, SD 57701, USA}
\affiliation{Dept. of Physics, University of Wisconsin, River Falls, WI 54022, USA}
\affiliation{Dept. of Physics and Astronomy, University of Rochester, Rochester, NY 14627, USA}
\affiliation{Department of Physics and Astronomy, University of Utah, Salt Lake City, UT 84112, USA}
\affiliation{Dept. of Physics, Chung-Ang University, Seoul 06974, Republic of Korea}
\affiliation{Oskar Klein Centre and Dept. of Physics, Stockholm University, SE-10691 Stockholm, Sweden}
\affiliation{Dept. of Physics and Astronomy, Stony Brook University, Stony Brook, NY 11794-3800, USA}
\affiliation{Dept. of Physics, Sungkyunkwan University, Suwon 16419, Republic of Korea}
\affiliation{Institute of Physics, Academia Sinica, Taipei, 11529, Taiwan}
\affiliation{Dept. of Physics and Astronomy, University of Alabama, Tuscaloosa, AL 35487, USA}
\affiliation{Dept. of Astronomy and Astrophysics, Pennsylvania State University, University Park, PA 16802, USA}
\affiliation{Dept. of Physics, Pennsylvania State University, University Park, PA 16802, USA}
\affiliation{Dept. of Physics and Astronomy, Uppsala University, Box 516, SE-75120 Uppsala, Sweden}
\affiliation{Dept. of Physics, University of Wuppertal, D-42119 Wuppertal, Germany}
\affiliation{Deutsches Elektronen-Synchrotron DESY, Platanenallee 6, D-15738 Zeuthen, Germany}

\author{R. Abbasi}
\affiliation{Department of Physics, Loyola University Chicago, Chicago, IL 60660, USA}
\author{M. Ackermann}
\affiliation{Deutsches Elektronen-Synchrotron DESY, Platanenallee 6, D-15738 Zeuthen, Germany}
\author{J. Adams}
\affiliation{Dept. of Physics and Astronomy, University of Canterbury, Private Bag 4800, Christchurch, New Zealand}
\author{S. K. Agarwalla}
\thanks{also at Institute of Physics, Sachivalaya Marg, Sainik School Post, Bhubaneswar 751005, India}
\affiliation{Dept. of Physics and Wisconsin IceCube Particle Astrophysics Center, University of Wisconsin{\textemdash}Madison, Madison, WI 53706, USA}
\author{J. A. Aguilar}
\affiliation{Universit{\'e} Libre de Bruxelles, Science Faculty CP230, B-1050 Brussels, Belgium}
\author{M. Ahlers}
\affiliation{Niels Bohr Institute, University of Copenhagen, DK-2100 Copenhagen, Denmark}
\author{J.M. Alameddine}
\affiliation{Dept. of Physics, TU Dortmund University, D-44221 Dortmund, Germany}
\author{S. Ali}
\affiliation{Dept. of Physics and Astronomy, University of Kansas, Lawrence, KS 66045, USA}
\author{N. M. Amin}
\affiliation{Bartol Research Institute and Dept. of Physics and Astronomy, University of Delaware, Newark, DE 19716, USA}
\author{K. Andeen}
\affiliation{Department of Physics, Marquette University, Milwaukee, WI 53201, USA}
% ...

\collaboration{IceCube Collaboration}
\thanks{analysis@icecube.wisc.edu}
\noaffiliation

\begin{abstract}
In this Letter, we present the results of a search for high-energy neutrinos produced by the annihilation of dark matter particles trapped in the Sun.
Using 9.3 and 10.4 years of data from the DeepCore and IceCube neutrino detectors, we establish world-best limits for spin-dependent interactions between dark matter and Standard Model particles for dark matter masses from tens of GeV to tens of TeV.
We additionally place constraints on the neutrino background produced by interactions of cosmic rays with the solar atmosphere.
\end{abstract}

\maketitle

\textbf{Introduction---}Solar neutrinos at \si{\MeV} energies have played a critical role in understanding both the Sun and particle physics more broadly.
These neutrinos confirmed that the Sun generates its power via nuclear fusion~\cite{Cleveland:1998nv,Davis:1968cp}, and the effort to understand the discrepancy between the theoretical predictions and observed data~\cite{Bahcall:1976zz} led to the discovery of flavor conversion during transport~\cite{SNO:2002tuh,Super-Kamiokande:1998kpq}, indicating that neutrinos have a non-zero mass and posing a significant challenge to the Standard Model (SM) of particle physics; however, there is reason to believe that solar neutrinos may have more to tell us at \si{GeV} energies.

In the SM, cosmic-ray interactions with solar nuclei create charged hadrons that ultimately decay into neutrinos~\cite{Seckel:1991ffa}.
This is the same process that produces neutrinos in Earth's atmosphere, and these neutrinos are analogously called \textit{solar atmospheric neutrinos}.
These neutrinos follow a well-predicted spectrum~\cite{Arguelles:2017eao,Ng:2017aur,Edsjo:2017kjk} that resembles that of their terrestrial counterparts but shifted to slightly higher energies since the charged hadrons lose less energy in the thinner solar atmosphere.
The gamma rays produced by the corresponding neutral hadrons have been observed in Fermi-LAT~\cite{Fermi-LAT:2011nwz,Ng:2015gya,Linden:2018exo} and HAWC~\cite{HAWC:2022khj} data.
Still, there are unexplained features in the spectrum~\cite{Tang:2018wqp}---including a dip around \SI{50}{GeV} and a hardening associated with the solar minimum---on which solar atmospheric neutrinos may help shed light.

Additionally, scenarios beyond the SM can give rise to high-energy solar neutrinos.
In this context, scenarios in which the dark matter (DM) that comprises roughly 84\% of the Universe's matter~\cite{Freese:2017idy,Planck:2018vyg} is made of fundamental particles~\cite{Bertone:2004pz} are of particular interest.
A global effort to detect such particles is underway~\cite{Liu:2017drf, Conrad:2017pms,Buchmueller:2017qhf,Arguelles:2019ouk}, and DM comprised of weakly interacting massive particles---particles that interact with SM particles at or below the weak scale---are particularly interesting.
Such DM would scatter on nuclei in the Sun, lose energy, and become gravitationally bound~\cite{Jungman:1995df}.
After further scatterings, an excess of DM would accumulate in the solar center.
Eventually, the DM would annihilate into SM particles, among which only neutrinos interact weakly enough to escape the dense center of the Sun.
This means we can search for indirect evidence of DM by looking for an excess of high-energy neutrinos from the Sun's direction.

\begin{figure*}[t]
    \centering
    \includegraphics[width=0.9\textwidth,
                  height=\textheight,
                  keepaspectratio]{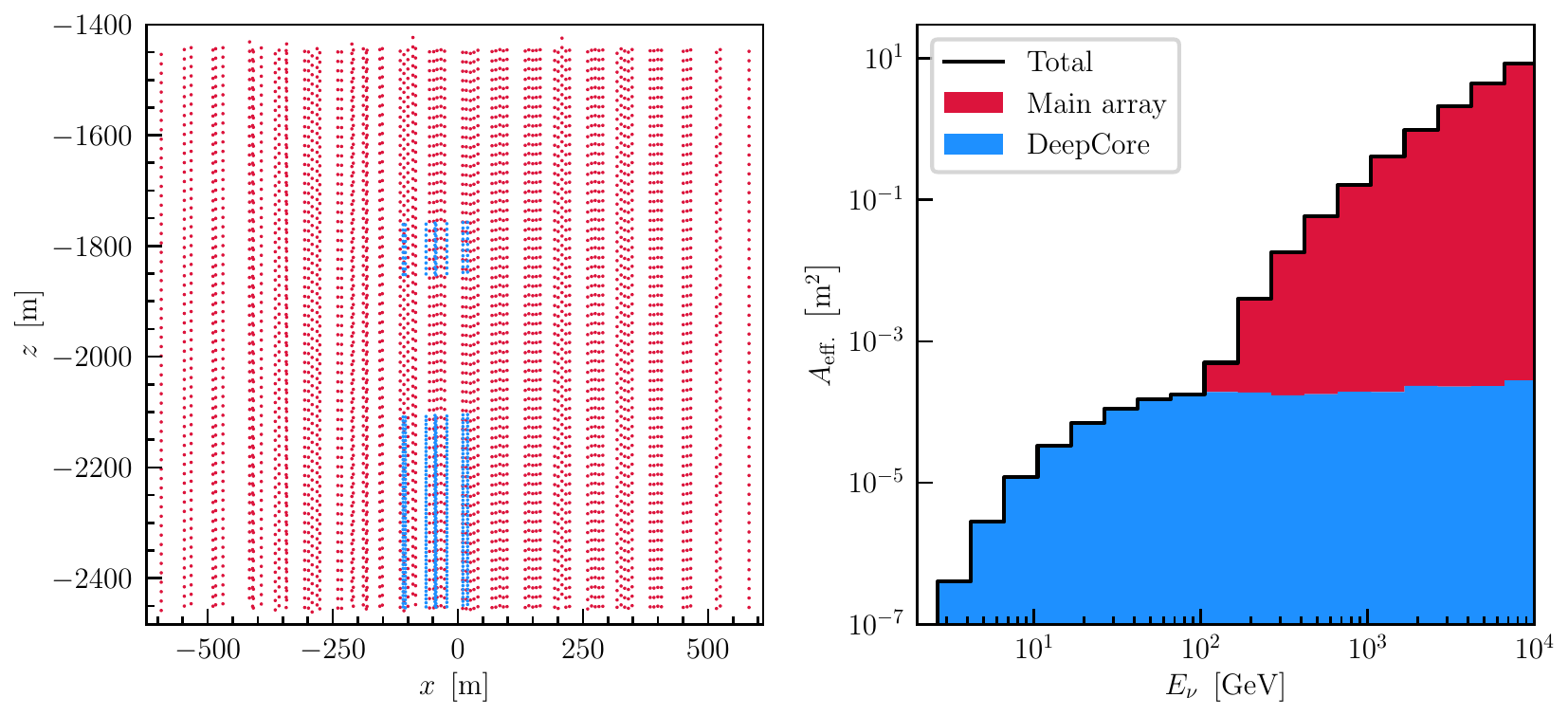}
    \caption{\textbf{\textit{Detector layout and effective area.}}
    The left-hand panel shows a view of the IceCube detector from the side.
    The main array is shown in red, while the smaller, denser DeepCore subarray is shown in blue.
    The right-hand panel shows the combined effective areas of both the full-detector and DeepCore-exclusive event selections.
    Each array's contribution is indicated by the color corresponding to the left-hand panel, while the black line represents the total effective area.
    The DeepCore selection utilizes the main array to veto events that are not fully contained within the sub-array and provides access to neutrinos with energies below 100 GeV.
    In contrast, the main array contributes the dominant portion above these energies.
    }
    \label{fig:detector}
\end{figure*}

Since the capture and annihilation rates are expected to be in equilibrium in the Sun~\cite{Jungman:1995df}, such searches are sensitive to the DM-proton scattering cross section.
Thus, this method complements terrestrial direct detection searches, which look for excess energy deposited via nuclear recoil on target nuclei~\cite{Liu:2017drf}.
Currently, the most sensitive direct detection experiments probe the mass range from \SI{1}{\GeV} to \SI{100}{\TeV}, using liquid xenon time-projection chambers~\cite{XENON:2018voc,LUX:2016ggv,PandaX-II:2020oim} or bubble chambers~\cite{PICO:2019vsc}.

The IceCube Neutrino Observatory (IceCube) is a gigaton-scale ice-Cherenkov detector near the geographic South Pole.
The 5,160 light-detecting digital optical modules (DOMs), arranged on 86 strings of 60 modules~\cite{IceCube:2016zyt}, detect the Cherenkov light produced by the charged byproducts of neutrino interactions.
In the main array, the DOMs have a vertical spacing of \SI{17}{\m} and interstring spacing of \SI{125}{\m}, while the denser DeepCore sub-array~\cite{IceCube:2011ucd} has a vertical spacing of \SI{7}{\m} and interstring spacing of $\approx\SI{70}{\m}$.
The left-hand panel of \cref{fig:detector} shows a side-on view of the detector with the low-energy DeepCore array in blue and the high-energy main array in red.
By combining these two arrays, IceCube can detect neutrinos with energies ranging from \SI{5}{GeV} to $\si{PeV}$.
Thus, IceCube can search for the soft spectrum of solar atmospheric neutrinos and probe a mass range of DM similar to that probed by direct detection experiments.

This Letter provides an update to and improvement upon IceCube's previous searches for solar atmospheric neutrinos~\cite{IceCube:2019ubb} and neutrinos from DM annihilation in the Sun~\cite{IceCube:2021xzo,IceCube:2016dgk}.
This analysis combines events from both a low-energy, DeepCore-based selection and events from a high-energy selection, using 9.3 years and 10.4 years of data, respectively.
In addition to this longer live-time, both selections use improved, machine-learning-based reconstruction techniques with improved angular and energy resolution, thus improving sensitivity to neutrinos from dark matter annihilation and cosmic-ray interactions in the solar atmosphere.
Additionally, this analysis employs an updated calculation of the neutrino yield from dark matter annihilation, which can significantly alter the spectrum for dark matter masses above the weak scale.

\textbf{Data Sample and Simulation---}This analysis combines the low-energy selection from IceCube's recent measurement of atmospheric neutrino oscillation parameters~\cite{IceCubeCollaboration:2024ssx} with the high-energy event selection used in the recent search for neutrino emission from astrophysical objects~\cite{IceCube:2022der}.
The only change to these selections is an additional cut on a low-level filter to remove any events from the low-energy selection that could occur in the high-energy selection to ensure the statistical independence between the samples.
The high-energy selection focuses on obtaining a high-purity selection of \numu{}\footnote{\numu{} refers to combined neutrinos and anti-neutrinos of the muon flavor. This convention will be maintained for all neutrino flavors throughout the text.} charged-current events.
Since the $\mu^{\pm}$ produced in this interaction can travel $\gtrapprox \SI{1}{\km}$ at energies $\gtrapprox \SI{300}{\GeV}$~\cite{Koehne:2013gpa}, these events leave long, track-like signatures in the detector.
This long lever arm allows relatively precise reconstruction of the $\mu^{\pm}$ direction.
In the energy regime targeted by the low-energy selection, the muon both produces less light and travels much shorter distances.
Additionally, this selection imposes containment cuts to ensure that the event only occurs within the DeepCore fiducial volume, limiting the track length of the outgoing $\mu^{\pm}$ from \numu{} charged-current interactions.
Thus, the clear advantage in angular resolution that \numu{} events have at higher energies is dramatically lessened, and this selection includes neutrinos of all flavors.
Quantitatively, the highest-energy \numu{} charged-current events have a median misreconstructed angle of $\SI{31}{\degree}$, while cascades have a reconstruction angle between \SI{58}{\degree} and \SI{38}{\degree}.
The right-hand panel of \cref{fig:detector} shows the effective area contribution as a function of energy for each selection in a color corresponding to the left-hand panel.
Furthermore, the total effective area, given by the sum over both selections, is shown as a black line.

In the energy range relevant to this analysis, i.e. \SI{5}{\GeV} to \SI{10}{\TeV}, the primary sources of events, and thus main backgrounds, are neutrinos and muons created when cosmic rays interact in Earth's atmosphere~\cite{Gaisser:2002jj}.
These are modeled using experimental data that has been directionally randomized.
Specifically, each event is assigned a random reconstructed right ascension.
Since the Sun is not near the poles of the equatorial coordinate system, this will tend to move signal events far from their true direction, thus eliminating the clustering effect of point source emission while maintaining the declination dependence of the atmospheric background rates.
This has the additional advantage that the analysis need not contend with systematics associated with the background estimation.

\begin{figure}
    \centering
    \includegraphics[width=0.95\linewidth,
                  height=\textheight,
                  keepaspectratio]
                  {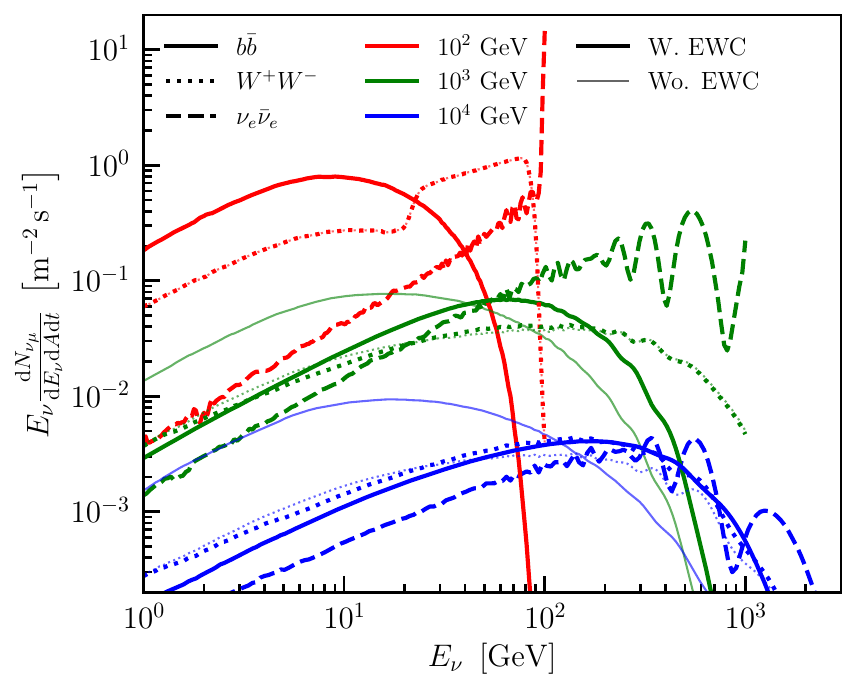}
    \caption{\textbf{\textit{Spectra of neutrinos for several annihilation channels and WIMP masses at Earth.}}
    The line styles represent a primary annihilation channel, while the colors represent different WIMP masses.
    If the annihilation hypothesis was previously studied in an IceCube analysis, a line with lower opacity shows the spectrum without including the EW correction.
    Note the large differences in the spectra from annihilation to $b\bar{b}$ at masses above the EW scale and the absence of such differences in the spectra from annihilation to $W^{+}W^{-}$.
    }
    \label{fig:initial_spectrum}
\end{figure}

This analysis uses a recent calculation~\cite{Arguelles:2017eao} of the flux of solar atmospheric neutrinos.
This calculation uses a hybrid model of the solar density profile based on~\cite{spruit:1974abc,Bahcall:2004pz,fontenla:2007abc,gingerich:1971abc}, and the \texttt{MCEq} cascade equation solver~\cite{Fedynitch:2015zma} to calculate the neutrino production from initial hadrons and secondary muons.
The resulting flux of neutrinos is then propagated from the Sun to the Earth's surface using the density matrix formalism as implemented in the \nusquids{} package~\cite{Arguelles:2021twb}.
The last-mile propagation from the Earth's surface to the detector is then performed with this same propagation framework.
Notably, this calculation computes the neutrino flux for different initial cosmic-ray flux and hadronic interaction models.
This analysis follows this calculation and takes the nominal spectrum as that resulting from the combined Gaisser-Honda with Hillas-Gaisser H4a primary cosmic-ray flux~\cite{Fedynitch_2012} and Sibyll-2.3c hadronic interaction model~\cite{Riehn:2015aqb,Riehn:2015oba} assuming a purely protonic atmosphere and MRS prompt model~\cite{Martin:2003us}.
Ref~\cite{Arguelles:2017eao} calculates the solar atmospheric flux with other interaction and cosmic-ray models; however, most fluxes differ primarily by a normalization, allowing for trivial conversion of limits.
Furthermore, the flux from the nominal calculation agrees with the recent calculation from Ref.~\cite{Ng:2017aur}, enabling simple comparison.
See \cref{sec:sanu} for further discussion.

Finally, to simulate the neutrino production from DM annihilation, this analysis uses the \charon{} package~\cite{Liu:2020ckq}.
It is assumed the DM annihilates into two monochromatic SM particles, and the prompt particle spectra from this process are read from tables.
These potentially unstable particles are then allowed to interact and decay until only neutrinos remain.
In this process, \charon{} uses a Monte Carlo approach to account for the competing effects of interaction and decay in the dense solar core.
After this step, the neutrino flux is propagated to the detector, again using the \nusquids{} package.

\charon{} has two distinct calculations of the initial particle tables, one computed using the \texttt{PYTHIA} 8.2 library~\cite{Sjostrand:2014zea} and one which includes a recent calculation of the electroweak (EW) effect~\cite{Bauer:2020jay}.
The EW effect becomes important when considering DM with masses above the EW scale, which may involve a decay to the unbroken SM.
Whereas previous methods for calculating the neutrino yield from DM annihilation either ignored~\cite{Niblaeus:2019gjk} or partially treated~\cite{Cirelli:2010xx}
this effect, \charon{} offers a full treatment.

\begin{figure*}[t]
    \centering
    \includegraphics[width=0.9\textwidth,
                  height=\textheight,
                  keepaspectratio]{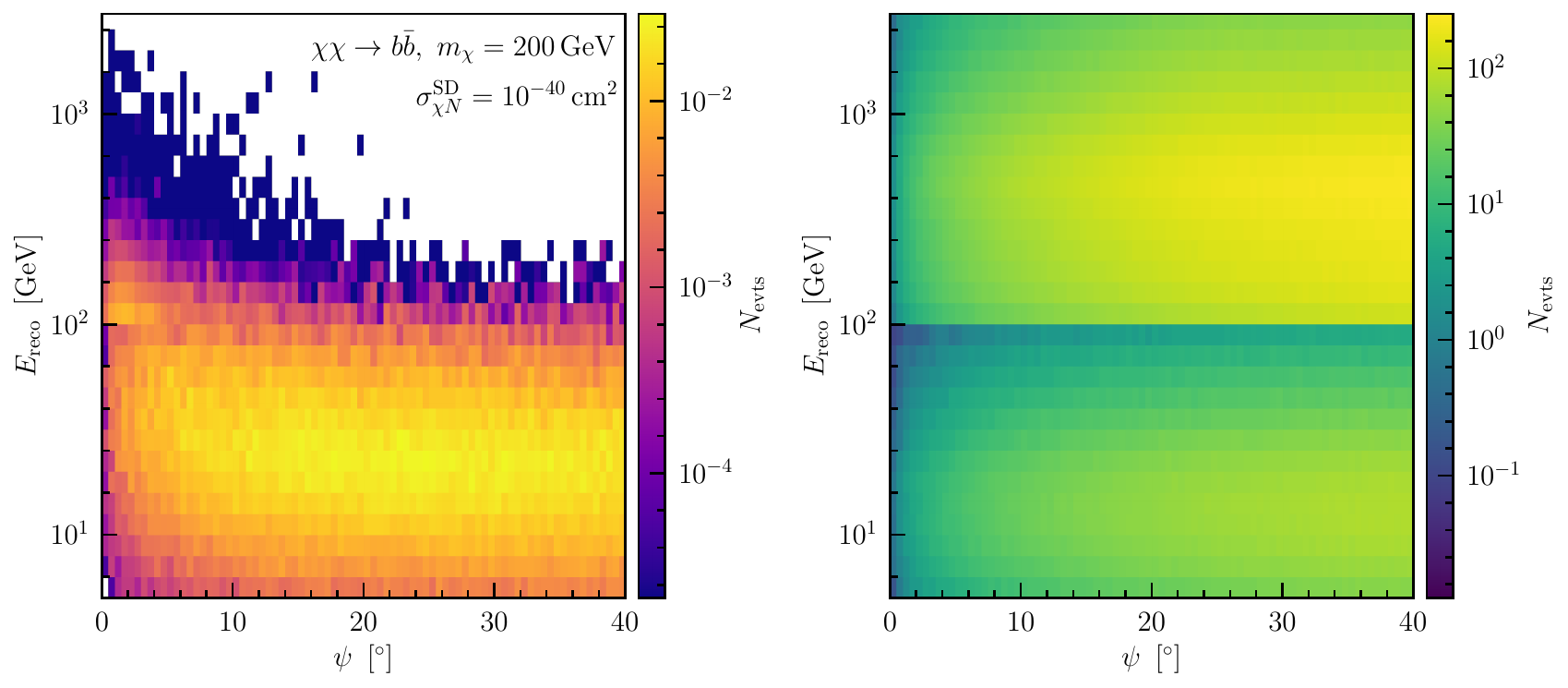}
    \caption{\textbf{\textit{Expected signal and background distributions as a function of the reconstructed energy,$E_{\mathrm{reco}}$ and reconstructed angular distance from the Sun, $\psi$.}}
    The left panel shows the expected distribution of events from 200 GeV WIMPs annihilating to $b\bar{b}$ as a function of the reconstructed energy and reconstructed angular distance from the Sun.
    The right panel shows the expected distribution of background events as a function of the same variables.
    The number of background events is larger, but the signal events are preferentially clustered towards the direction of the Sun.
    The line visible at 100 GeV is the interface between the two event selections, and the relative deficit of events near the Sun in the background distribution is due to the diminishing phase space.
    }
    \label{fig:distributions}
\end{figure*}

Including this effect leads to dramatic changes in the initial neutrino spectra for certain channels.
\cref{fig:initial_spectrum} compares the neutrino spectra with and without this effect after propagation to the Earth's surface for three dark-matter masses and three annihilation channels.
The lowest mass, \SI{100}{\GeV} is below the EW scale, so the fluxes match identically for all channels.
For DM with masses of \SI{1}{\TeV} and \SI{10}{\TeV}, a large difference emerges for annihilation to a $b\bar{b}$ pair.
This arises from the prompt emission of a weak gauge boson decaying to a hard neutrino, rather than all neutrinos coming from the soft byproducts of hadronization.
Additionally, there is a smaller shift in the spectra from annihilation to a $W^{+}W^{-}$ pair as this new emission channel spreads the energy across more weak bosons, leading to lower energy neutrinos.

\textbf{Analysis Methodology---}For analysis, the data is binned in reconstructed quantities to obtain expected distributions for background, solar atmospheric neutrinos, and neutrinos from dark matter annihilation.
The data in the high-energy selection is binned in reconstructed energy and reconstructed angular distance from the center of the Sun.
The data in the low-energy selection is additionally binned in a morphological classification variable meant to differentiate events that originate from \numu charged-current interactions from other events.
This is a useful quantity because neutrinos from the Sun will be fully mixed, while for sufficiently high-energies---$\gtrapprox\SI{40}{\GeV}$---the atmospheric neutrinos will maintain a preference for \numu.
Thus, an excess of non-\numu events at specific energies can help discriminate signal and background.

The Monte-Carlo-derived distributions---i.e. for solar atmospheric neutrinos and neutrinos from DM annihilation---are constructed by selecting a time within the analysis window, calculating the Sun's position, and discarding any events outside of the Sun's physical extent.
This procedure is then repeated many times to derive a smooth event distribution.
In addition to computing this for the solar atmospheric flux, this is derived for 51 DM hypothesis with masses between \SI{20}{\GeV} and \SI{10}{\TeV} annihilating to $b\bar{b}$, $W^{+}W^{-}$, $\tau^{+}\tau^{-}$, and $\nu_{\alpha}\bar{\nu}_{\alpha}$ pairs.
The left panel of \cref{fig:distributions} shows the expected number of events for a reference DM hypothesis at a reference cross section of $\sigma_{\chi p}=10^{-40}\,\si{\cm\squared}$ in both event selections; the low-energy distribution has been summed over the morphological discriminator to facilitate this display.
The high-energy selection is the dominant contribution at energies above \SI{100}{\GeV}, as evidenced by events clustering in the \SI{3}{\degree} nearest the Sun.
Below this energy, where angular reconstruction is challenging, the events are spread over a much larger angular range.
The small discontinuity at \SI{100}{\GeV} is caused by a cut on the reconstructed energy in the high-energy selection.

The data-derived background distribution is computed by taking all events that pass the selections and assigning a random right ascension drawn uniformly from the distribution $\left[0, 2\pi \right)$ and binning in the appropriate variables.
Since there are typically not enough events to fill the bins with appropriate sample size, this procedure is carried out tens of times and the resulting distributions are averaged.
The right panel of \cref{fig:distributions} shows the background distribution for both event selections.
As in the previous case, the low-energy selection has been summed over the morphology discriminator.
In contrast to the signal distributions, the number of background events grows towards larger angles due to the growing phase space.

This analysis then uses a binned Poisson likelihood~\cite{ParticleDataGroup:2024cfk} given by:
\begin{equation*}
    \lh(n | \theta) = \prod_{i}\frac{e^{-\mu_{i}\left(\theta\right)}\mu_{i}\left(\theta\right)^{n_{i}}}{n_{i}!}, 
\end{equation*}
where $\theta=\left(\alpha^{\chi},\,\alpha^{\mathrm{sa}},\, \alpha^{\mathrm{a}}\right)$ is the model parameters, i.e. the DM, solar atmospheric, and conventional atmospheric normalizations with respect to the nominal; $\mu\left(\theta\right)$ is the expected number of events in a particular model; $n$ is the observed number of events; and $i$ indexes the bin.
In order to test the DM hypothesis, the test statistic is given by twice the log-likelihood ratio between the best fit with and without a contribution from DM:
\begin{align*}
    \mathrm{TS} = -2\left[\ln(\lh(n | 0,\, \hat{\alpha}^{\mathrm{sa}},\, \hat{\alpha}^{\mathrm{a}})) - \ln(\lh(n|\check{\alpha}^{\chi},\, \check{\alpha}^{\mathrm{sa}},\, \check{\alpha}^{\mathrm{a}}))\right].
\end{align*}
Here, $\hat{\alpha}^{\mathrm{y}}$ and $\check{\alpha}^{\mathrm{y}}$ denote the best-fit normalization in the background-only and signal-plus-background models, respectively.
This framework is then adapted to search for solar atmospheric neutrinos by setting $\alpha^{\chi}=0$ and fitting with and without the contribution from solar atmospheric neutrinos, i.e.:
\begin{align*}
    \mathrm{TS} = -2\left[\ln(\lh(n|0,\, 0,\, \hat{\alpha}^{\mathrm{a}})) - \ln(\lh(n|0,\, \check{\alpha}^{\mathrm{sa}},\, \check{\alpha}^{\mathrm{a}}))\right].
\end{align*}
In both the DM and solar atmospheric cases, the background-only test-statistic distributions follow a modified $\chi^{2}$ distribution with one degree of freedom.
The modification arises from allowing only positive normalizations to be fit, leading to 50\% of events with a test statistic equal to zero.

\begin{figure}
    \centering
    \includegraphics[width=0.5\textwidth,
                  height=\textheight,
                  keepaspectratio]{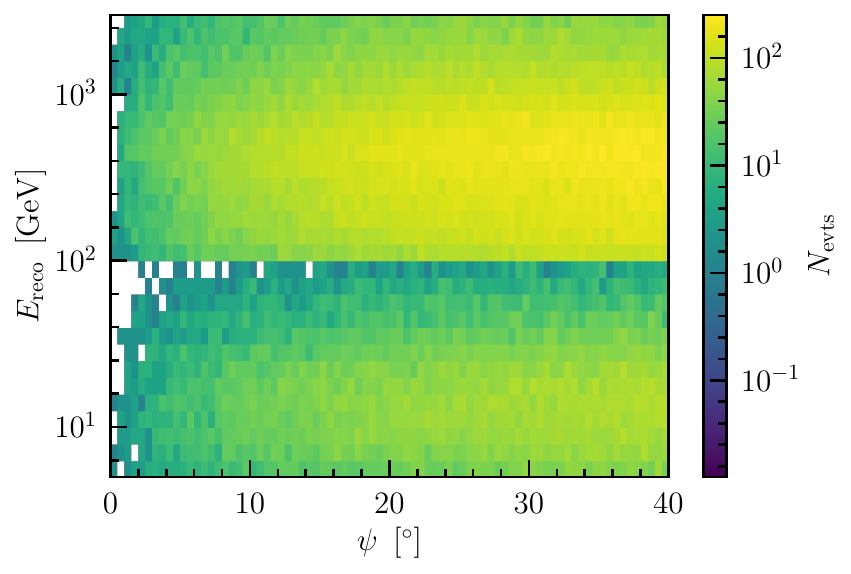}
    \caption{\textbf{\textit{Unblinded analysis data.}}
    No excess of neutrinos from the direction of the Sun was observed.
    The most significant DM hypothesis was annihilation to $b
    \bar{b}$ for $m_{
    \chi
    }=\SI{20}{\GeV}$, with a local significance of $1.76
    \sigma$.}
    \label{fig:data}
\end{figure}

\textbf{Results---}The observed data binned in angular distance from the Sun and reconstructed energy are shown in \cref{fig:data}.
Of the 51 DM hypotheses this analysis considered, 32 had a best-fit value at no DM contribution to the flux.
Of the nineteen DM hypotheses that fit a non-zero contribution, only DM with a mass of \SI{20}{\GeV} annihilating to a $b\bar{b}$ pair preferred the DM-plus-SM hypothesis with a $p$-value $< 0.1$ and had a $p$-value of 0.079 or $1.76\sigma$.
Given the relatively low significance and high number of tested hypotheses, we do not report a significant neutrino excess from DM annihilation in the Sun.

\begin{figure}
    \centering
    \includegraphics[width=0.48\textwidth,
                  height=\textheight,
                  keepaspectratio]{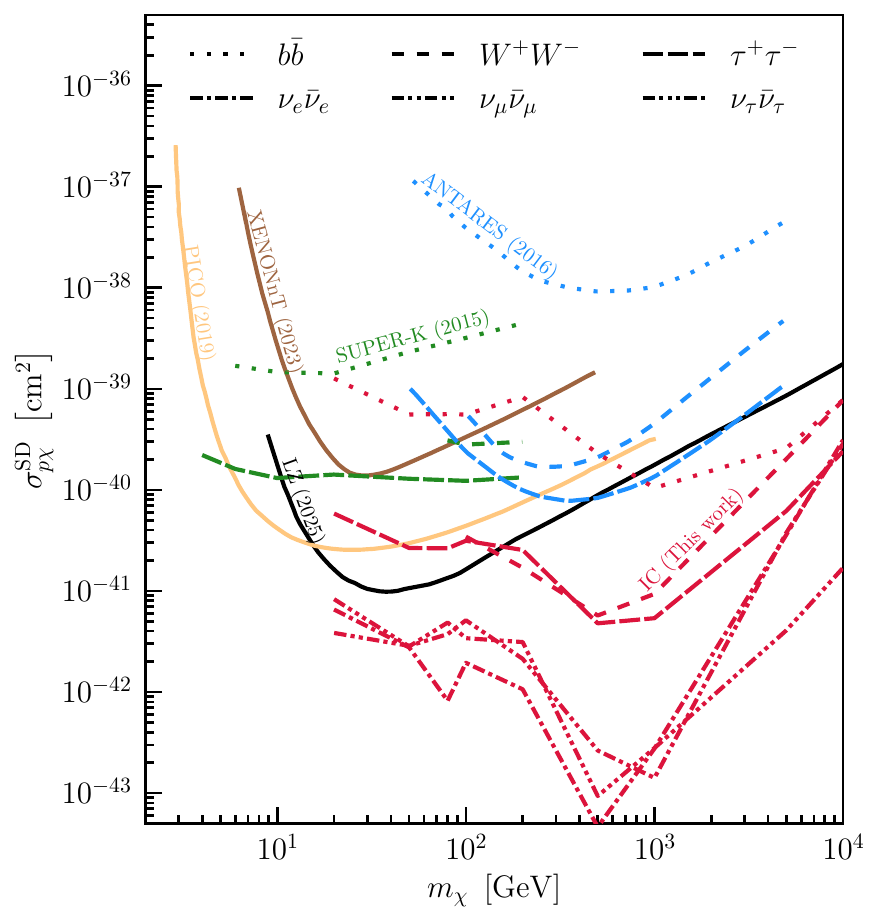}
    \caption{\textbf{\textit{Limits on the spin-dependent DM-proton cross section.}}
    Since no excess of neutrinos was observed from the direction of the Sun, limits on the spin-dependent WIMP-proton cross section are obtained.
    These new limits are compared to existing limits on this cross section from other experiments.
    Direct detection experiments are shown with solid lines, whereas indirect detection experiments are shown with line styles corresponding to the annihilation channel.
    This analysis achieves world-leading limits for most annihilation channels for WIMPs above \SI{200}{\GeV}.
    Additionally, this analysis improves on the high-mass $b\bar{b}$ limits by more than an order of magnitude, highlighting the importance of the electroweak correction for hadronic annihilation channels.
    These limits are compared to the previous limits from PICO~\cite{PICO:2019vsc}, ANTARES~\cite{ANTARES:2016xuh}, XENONnT~\cite{XENON:2023cxc}, LUX-ZEPLIN~\cite{LZ:2024zvo}, and Super-Kamiokande~\cite{Super-Kamiokande:2015xms}.
}
    \label{fig:limits}
\end{figure}

We set limits at the 90\% confidence level by increasing the dark matter contribution until the test-statistic difference relative to the best-fit point reaches 1.64, which corresponds to the 90\% containment of the modified $\chi^{2}$ distribution discussed above~\cite{Cowan:2010js}.
The limits on the spin-dependent DM-proton cross section for this analysis are compared to existing limits in the right panel of \cref{fig:limits}.

There are a few notable features in these limits.
In the cases of annihilation to $W^{+}W^{-}$, $\tau^{+}\tau^{-}$, and $\nu_{\alpha}\bar{\nu}_{\alpha}$ pairs, this analysis explores yet-unprobed parameter space for DM with a mass above \SI{200}{\GeV}.
Furthermore, this analysis dramatically enhances limits on annihilation to a $b\bar{b}$ pair at high masses, improving upon the previous IceCube limits~\cite{IceCube:2016dgk} by a factor of $\sim40$ and the ANTARES limits~\cite{ANTARES:2016xuh} by a factor of $\sim100$.
Finally, this analysis sets the most stringent indirect detection constraints in all considered DM hypotheses except DM with a mass of \SI{20}{\GeV} annihilating to a $b\bar{b}$ pair.
These limits on the DM-proton cross section can be recast as limits on other DM scattering cross sections, namely the spin-dependent DM-proton cross section and the DM-electron cross section.
This analysis sets world-leading constraints on the latter cross section, while the constraints on the former are competitive with those of direct detection experiments for certain annihilation channels.
This is discussed further in \cref{sec:other_limits}.

The fit that only considered a contribution from SM sources, i.e., conventional and solar atmospheric neutrinos, found a best-fit value of no contribution from solar atmospheric neutrinos.
By using the same methods, this analysis set a limit on the flux of solar atmospheric neutrinos at 2.48 times the nominal normalization.
This is in line with expectations since sensitivity studies showed that the nominal flux would have needed to be 2.78 times larger in order to be observed 50\% of the times.

\textbf{Conclusions---}This analysis significantly improves over previous limits from IceCube and sets world-leading constraints on the DM-proton scattering cross section for DM masses above \SI{100}{\GeV}.
Furthermore, this analysis sets the most stringent indirect detection constraints on all except one of the considered DM hypotheses.
While the constraints for masses above \SI{100}{\GeV} should stand alone until other gigaton-scale neutrino telescopes~\cite{KM3Net:2016zxf,Avrorin:2015wba,P-ONE:2020ljt,TRIDENT:2022hql} are constructed and have taken comparable data, limits on DM below this mass range will be improved by the forthcoming IceCube Upgrade~\cite{IceCube:2023ins} and KM3NeT-ORCA~\cite{KM3Net:2016zxf} detectors.
This possibility is especially interesting given that the most significant DM hypothesis in this analysis was found at the lowest considered mass.
Furthermore, this analysis sets a limit on neutrinos from cosmic-ray interactions at 2.48 times the nominal model, suggesting novel approaches or significantly more data may be necessary to observe these neutrinos.

\textbf{Acknowledgements---}The IceCube collaboration acknowledges the significant contributions to this manuscript from J.~Lazar, Q.~R.~Liu, and I.~Mart\'{i}nez-Soler.
We acknowledge the support from the following agencies: USA {\textendash} U.S. National Science Foundation-Office of Polar Programs,
U.S. National Science Foundation-Physics Division,
U.S. National Science Foundation-EPSCoR,
U.S. National Science Foundation-Office of Advanced Cyberinfrastructure,
Wisconsin Alumni Research Foundation,
Center for High Throughput Computing (CHTC) at the University of Wisconsin{\textendash}Madison,
Open Science Grid (OSG),
Partnership to Advance Throughput Computing (PATh),
Advanced Cyberinfrastructure Coordination Ecosystem: Services {\&} Support (ACCESS),
Frontera computing project at the Texas Advanced Computing Center,
U.S. Department of Energy-National Energy Research Scientific Computing Center,
Particle astrophysics research computing center at the University of Maryland,
Institute for Cyber-Enabled Research at Michigan State University,
Astroparticle physics computational facility at Marquette University,
NVIDIA Corporation,
and Google Cloud Platform;
Belgium {\textendash} Funds for Scientific Research (FRS-FNRS and FWO),
FWO Odysseus and Big Science programmes,
and Belgian Federal Science Policy Office (Belspo);
Germany {\textendash} Bundesministerium f{\"u}r Bildung und Forschung (BMBF),
Deutsche Forschungsgemeinschaft (DFG),
Helmholtz Alliance for Astroparticle Physics (HAP),
Initiative and Networking Fund of the Helmholtz Association,
Deutsches Elektronen Synchrotron (DESY),
and High Performance Computing cluster of the RWTH Aachen;
Sweden {\textendash} Swedish Research Council,
Swedish Polar Research Secretariat,
Swedish National Infrastructure for Computing (SNIC),
and Knut and Alice Wallenberg Foundation;
European Union {\textendash} EGI Advanced Computing for research, and Horizon 2020 Marie Sk\l{}odowska-Curie Actions;
Australia {\textendash} Australian Research Council;
Canada {\textendash} Natural Sciences and Engineering Research Council of Canada,
Calcul Qu{\'e}bec, Compute Ontario, Canada Foundation for Innovation, WestGrid, and Digital Research Alliance of Canada;
Denmark {\textendash} Villum Fonden, Carlsberg Foundation, and European Commission;
New Zealand {\textendash} Marsden Fund;
Japan {\textendash} Japan Society for Promotion of Science (JSPS)
and Institute for Global Prominent Research (IGPR) of Chiba University;
Korea {\textendash} National Research Foundation of Korea (NRF);
Switzerland {\textendash} Swiss National Science Foundation (SNSF).

\appendix

\section{End Matter}

\subsection*{Nominal Solar Atmospheric Neutrino Model}
\label{sec:sanu}

To calculate IceCube's sensitivity to the solar atmospheric neutrinos, the present work uses a recent calculation by Arg\"{u}elles \textit{et al.}~\cite{Arguelles:2017eao}.
This calculation uses the \texttt{MCEq} software to compute the neutrino yield from primary hadrons and secondary muons resulting from cosmic-ray interactions as a function of the impact parameter.
The final neutrino flux is then integrated over the radius of the Sun, and the emerging neutrino flux is propagated to Earth's surface.
The propagation is averaged over a whole year to account for the non-constant Earth-Sun distance.
The year-averaged flux is then propagated to the detector using the same propagation framework used for the Earth-Sun propagation.

This work computes the flux for several primary cosmic-ray fluxes and hadronic interaction models.
One reason the nominal solar atmospheric model---that arises from the combined Gaisser-Honda with Hillas-Gaisser H4a primary cosmic-ray flux~\cite{Fedynitch_2012} and Sibyll-2.3c hadronic interaction model~\cite{Riehn:2015aqb,Riehn:2015oba}---was chosen is that it is an optimistic scenario.
Since the different models differ primarily by a normalization, limits on one model can be converted to limits on another.

\begin{figure}[t]
    \centering
    \includegraphics[width=0.95
    \linewidth]{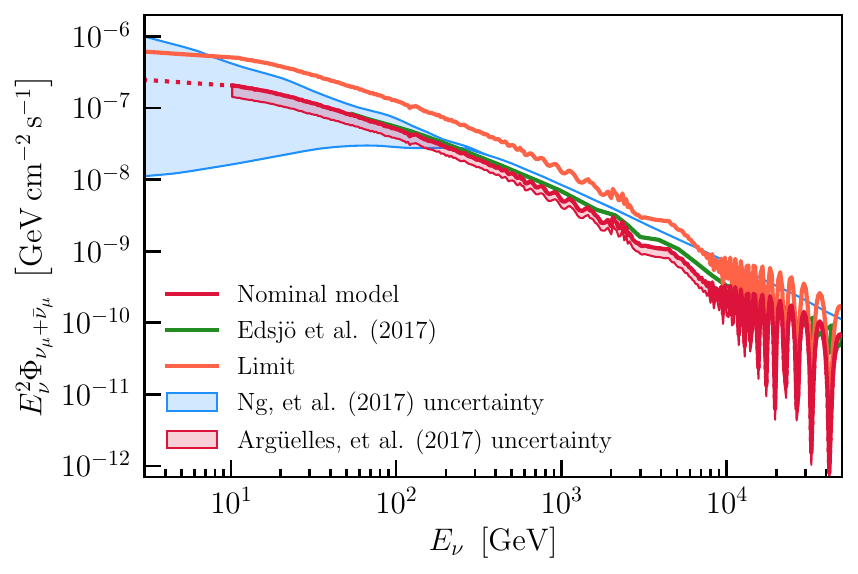}
    \caption{\textbf{\textit{Comparison of solar atmospheric fluxes.}}
    The nominal model is shown as a thick red line.
    This model had to be extrapolated beyond the limits of the initial calculation between \SI{1}{\GeV} and \SI{10}{\GeV}.
    This was done using a quartic spline, and the result is shown as a dashed red line.
    The uncertainty region from the Arg\"{u}elles \textit{et al.}~\cite{Arguelles:2017eao} work---resulting from the different primary cosmic-ray fluxes and hadronic interaction models---is shown as the shaded red region.
    The uncertainty region from Ng \textit{et al.}~\cite{Ng:2017aur} results from uncertainties on the impact of the solar magnetic field on the primary cosmic-ray flux.
    The coral line shows the limit set by this analysis at 2.48 times the nominal model.
    }
    \label{fig:sanu_comparison}
\end{figure}

In addition to the nominal calculations, two other studies of solar atmospheric neutrinos were published nearly simultaneously.
The work by Ng \textit{et al.}~\cite{Ng:2017aur} focused primarily on the implications of solar atmospheric neutrinos on the sensitivity floor for indirect DM searches due to these neutrinos and did not recompute a solar atmospheric spectrum.
The work by Edsj\"{o} \textit{et al.}~\cite{Edsjo:2017kjk}, performs a similar calculation to that of the nominal model, albeit within a different framework.
The calculated flux agrees relatively well with the nominal flux from this work.
Thus, the limits obtained in this work should be comparable to limits on the Edsj\"{o} \textit{et al.} flux, highlighting another reason for selecting the nominal model.
See \cref{fig:sanu_comparison} for a comparison of the solar atmospheric fluxes in these recent works.

\subsection*{Limits on Spin-Independent and Electron-DM Scatterings}
\label{sec:other_limits}

\begin{figure}[ht]
    \centering
    \includegraphics[width=0.95\linewidth]{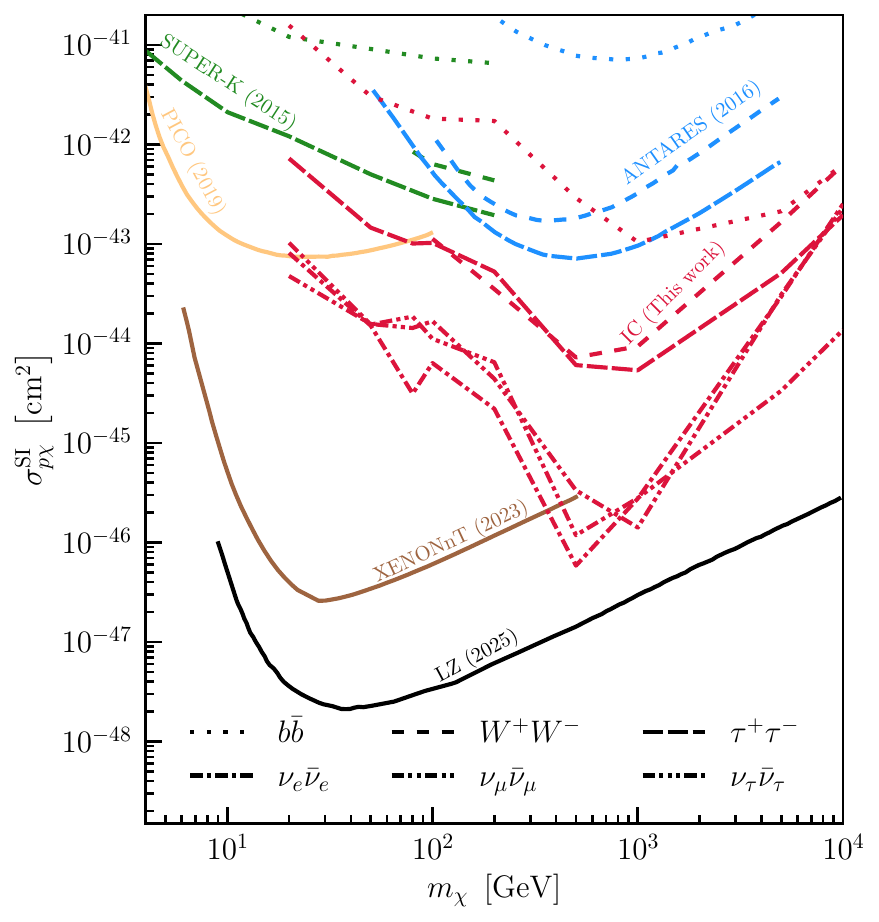}
    \caption{\textbf{\textit{Limits on the spin-independent DM-proton cross section.}}
    These new limits are compared to existing limits on this cross section from other experiments.
    Direct detection experiments are shown with solid lines, whereas indirect detection experiments are shown with line styles corresponding to the annihilation channel.
    These limits are compared to the previous limits from PICO~\cite{PICO:2019vsc}, ANTARES~\cite{ANTARES:2016xuh}, XENONnT~\cite{XENON:2023cxc}, LUX-ZEPLIN~\cite{LZ:2024zvo}, and Super-Kamiokande~\cite{Super-Kamiokande:2015xms}
    }
    \label{fig:si_limits}
\end{figure}

The limits shown in the main text can be reinterpreted as limits on different scattering cross sections by rescaling by the ratio of the annihilation rates for each cross section.
Under the assumption that the distribution in the Sun has equilibrated, this is given by the ratio of the capture rates.
In this section, we present limits on the spin-independent cross section in \cref{fig:si_limits} and compare them to the limits on this same quantity from other direct- and indirect-detection experiments.
As in the main text, we use the calculation in Ref~\cite{Jungman:1995df} to compute the capture rate for this cross section.

\begin{figure}[ht]
    \centering
    \includegraphics[width=0.99\linewidth]{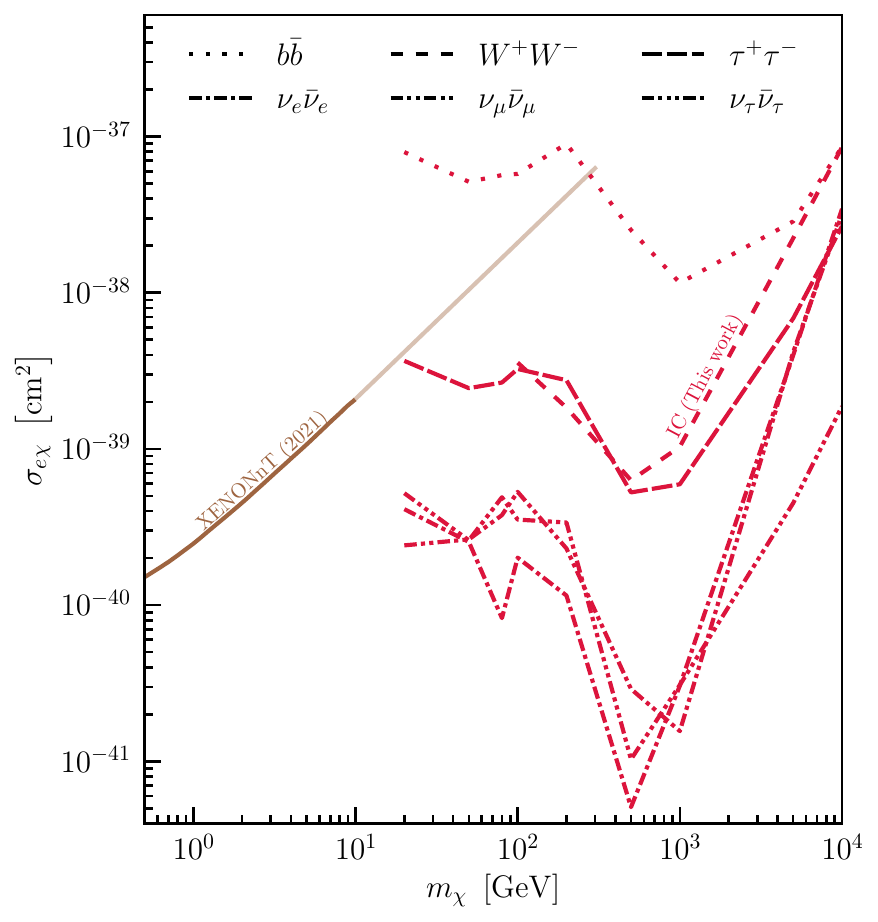}
    \caption{\textbf{\textit{Limits on the DM-electron cross section.}}
    These new limits are compared to existing limits on this cross section from other experiments.
    Direct detection experiments are shown with solid lines, whereas indirect detection experiments are shown with linestyles corresponding to the annihilation channel.
    These limits are compared to results from the XENON1T~\cite{XENON:2019gfn}.
    Results scaled according to $m_{\chi}$ are shown as a lighter line for comparison at masses higher than had been reported in the original work.
    }
    \label{fig:electron_limits}
\end{figure}

Additionally, we show the limits on the DM-electron cross section in \cref{fig:electron_limits}.
The capture rate has been computed using the calculation from Ref.~\cite{Garani:2017jcj}.
As this figure shows, this phase space has been much less explored.
The limits from this work are the first collaboration results in the mass range this analysis explores.
It is worth noting that there have been several theoretical reinterpretations of previous results in Refs.~\cite{Maity:2023rez,Nguyen:2025ygc,Krishna:2025ncv}.

\FloatBarrier
\bibliographystyle{apsrev4-1}
\bibliography{main}

\end{document}